\documentclass  {article}
\begin{document}

\title{%
Planck-scale relativity from quantum $\kappa$-Poincar\'e algebra}
\author{ J.\ Kowalski--Glikman\thanks{e-mail
address jurekk@ift.uni.wroc.pl}\\ Institute for Theoretical
Physics\\ University of Wroc\l{}aw\\ Pl.\ Maxa Borna 9\\
Pl--50-204 Wroc\l{}aw, Poland} \maketitle

\begin{abstract}
Extending the commutator algebra of quantum $\kappa$-Poincar\'e symmetry to the whole of the phase space, and assuming that this algebra is to be covariant under action of deformed Lorentz generators, we derive the transformation properties of positions under the action of deformed boosts. It turns out that these transformations leave invariant the quadratic form in the position space, which is the Minkowski metric and that the boosts saturate. The issues of massless and massive particles motion as well as time dilatation and length contraction in this new framework are also studied.
\end{abstract}
\clearpage

\section{Introduction}

In the recent years the $\kappa$-Poincar\'e symmetry algebra \cite{lunoruto, maru, luruto, luruza, luno} being initially an abstract mathematical construction has slowly started attract interest of physicists, trying to investigate its prediction in various physical contexts. In particular it was used in the analysis of anomalies observed in the spectrum of cosmic rays and  high energy gamma rays \cite{gacpir} (where the results obtained indicate that the parameter $\kappa$ in the algebra below should be of order of Planck mass), consistency of the inflationary fluctuations satisfying the $\kappa$-Poincar\'e dispersion relation with the observed Harrison-Zel'dovich spectrum \cite{jkgcosm}, and the possible role played by this algebra in the varying speed of light cosmology \cite{stjo}. Recently   a number of papers appear \cite{gac1}, \cite{gac2}, \cite{jkgminl} (see also \cite{gacnew} for up-to-date review), in which the proposals has been put forward concerning a {\em kinematical}\footnote{The world ``kinematical'' should be stressed here, as it indicates that the theory which is to be developed here differs both conceptually and technically from other proposals concerning breaking of Lorentz symmetry (see, e.g., \cite{kost}, \cite{jac}).} structure of space  and time at the Planck scale, based on the $\kappa$-Poincar\'e  algebra that may replace the standard special-relativistic Lorentz--Einstein--Minkowski structure. The central postulate of this new, so-called        ``doubly special special relativity'' was  an appearance of the second  observer-independent length (or mass) scale, in addition to the special relativistic, universal and observer-independent velocity scale, given by the speed of light.

Of course, the main motivation for research that has been undertaken in \cite{gac1}, \cite{gac2}, \cite{jkgminl}, whose direct continuation is the present paper, is the hope that they may shed some light on the problem of what is quantum theory of gravity. Usually in both string and loop approach to quantum gravity it is implicitly assumed that the kinematic structure of space and time, given by special relativity, is a correct starting point to construct a quantum theory of gravity (though, of course these approaches differ in a way in which this implicit assumption is implemented). Here, following the seminal papers by Giovanni Amelino-Camelia \cite{gac1}, \cite{gac2} I take a different view, namely that perhaps on our way leading to quantum gravity we must drastically revise already our concepts concerning physics of special relativity. This view is supported additionally by the fact that such a revision may provide us with explanation of already observed anomalies in cosmic ray physics,  and that one will be able to confront the prediction of such a revised theory with new experimental data already in a near future (see \cite{gacpir} and \cite{gacnew}).

Unfortunately, most of  papers dealt devoted to ``$\kappa$-Poincar\'e physics'' dealt only with the momentum sector of the  phase space, and the space-time transformations were not analyzed (see however \cite{maru}), the reason being that there apparently does not exist a simple guideline, telling how the momentum algebra is to be extended to the whole of the phase space. In this paper I will  present a proposal how this can be done and investigate its physical consequences.

\section{Preliminaries}

Let us start with the $\kappa$-Poincare algebra
\begin{eqnarray} \label{1}
  [M_{\mu\nu},M_{\rho\tau}]
=&&
\displaystyle
\left(\eta_{\mu\tau}M_{\nu\rho} -
\eta_{\nu\rho}M_{\nu\tau} +\eta_{\nu\rho}M_{\mu\tau}
  -\  \eta_{\nu\tau}M_{\mu\rho} \right),
\cr \cr
 [M_{i},P_{j}] =&&  \epsilon_{ijk}P_{k}\, ,
\quad [M_{i},P_{0}]= 0\, ,
\cr \cr
\displaystyle
  [N_{i}, P_{j}] = && \delta_{ij}
 \left( {\kappa\over 2} \left(
 1 -e^{-2{P_{0}/ \kappa}}
\right) + {1\over 2\kappa} \vec{P}\, ^{ 2}\, \right)
- \ {1\over \kappa} P_{i}P_{j} ,
\cr \cr
 \left[N_{i},P_{0}\right] = && P_{i}\, ,
\cr \cr
\displaystyle
  [P_{\mu},P_{\nu}] = && 0\, ,
\end{eqnarray}
where $P_\mu =(P_i, P_0)$ are space and time components of
four-momentum, and $M_{\mu\nu}$ are deformed Lorentz generators consisting of
rotations $M_k = \frac12\epsilon_{ijk}M_{ij} $ and boosts
$N_i = M_{0i}$. Moreover, $\kappa$ is a scale parameter, usually identified with the Planck mass (I use the units in which the universal speed of light, $c=1$). Let us note that this parameter could be either positive or negative, leading to completely different physical picture. This can be seen from the form of the Casimir operator of this algebra, to wit

\begin{equation}\label{2}
  {\cal M}^2 = \left(2\kappa \sinh\left(\frac{P_0}{2\kappa}\right)\right)^2 - \vec{P}\,{}^2 e^{P_0/\kappa}.
\end{equation}
Consider a massless particle, $M=0$. For $\kappa$ negative, both $|\vec{P}|$ and $P_0$ have infinite range, and the energy-dependent speed of light defined to be

\begin{equation}\label{3}
 {\cal C} = \frac{d\, P_0}{d\, |\vec{P}|} = \frac{1}{1 - |\vec{P}|/\kappa}
\end{equation}
goes to zero, when energy tends to infinity. In the case when $\kappa$ is positive, there is a maximal, observer-independent momentum $|\vec{P}| =\kappa$, which is reached at infinite energy and the corresponding speed of light (\ref{3}) is also infinite (see \cite{jkgminl, rbgacjkg} for details). This observation provides another important motivation for the exercise presented in this paper. Namely, the discovery of the observer-independent mass scale  was based solely on the analysis of the momentum sector, characterized by a single constant $\kappa$. It is therefore of extreme interest to understand the position sector of the phase space, which is necessary e.g., to study the particles motion, investigate the structure of space-time, etc.

To achieve this goal let us note that the algebra (\ref{1}) is only a part of the whole {\em quantum} algebra structure of $\kappa$-Poincar\'e. Another part consists of the so-called co-product, which is in  direct relation with non-commutativity of space and time, leading to the following commutators  \cite{szak}, \cite{maru}

\begin{equation}\label{4}
 [ X_0, X_i ] = i\ell\, X_i, \quad [ X_i, X_j ] =0,
\end{equation}
where $\ell$ is a parameter of dimension of length. The simplest way to check the consistency of (\ref{4}) is as follows. Take these equation, find the corresponding star product and apply it to form a product of two plane waves. This gives an expression of (deformed) sum of momenta which is to be identical with an expression obtained from the co-product of the quantum $\kappa$-Poincar\'e algebra. In this way one also can find out what  the value of the parameter $\ell$ is:   it turns out that

\begin{equation}\label{rel}
  \ell = - \hbar/\kappa,
\end{equation}
  and this condition is equivalent to the internal self-consistency of the {\em quantum} $\kappa$-Poincar\'e algebra.  

Now the  goal would be to extend the commutator algebra (\ref{1}), \ref{4}) to an algebra defined on the whole phase space. To do that I demand that the resulting algebra, as a whole,  be {\em covariant} with respect to the deformed Lorentz transformation, described by (\ref{1}), by which I mean that right and  left hand sides transform under deformed Lorentz transformations in the same way (which is, of course, equivalent to Jacobi identities being satisfied.) Since the construction I am to present below will be manifestly rotationally invariant, I will concentrate on checking the invariance with respect to boosts in direction $i$,
\begin{eqnarray} \label{5}
\displaystyle
\delta_i(P_j) \equiv  [N_{i}, P_{j}] = && \delta_{ij}
 \left( {\kappa\over 2} \left(
 1 -e^{-2{P_{0}/ \kappa}}
\right) + {1\over 2\kappa} \vec{P}\, ^{ 2}\, \right)
- \ {1\over \kappa} P_{i}P_{j} ,
\cr \cr
\delta_i(P_0) \equiv  \left[N_{i},P_{0}\right] = && P_{i}\, ,
\end{eqnarray}

\section{The   Poisson bracket algebra}

Before turning to the computation I will replace quantum commutators with Poisson brackets. The reason for  doing so is twofold. First of all it makes it possible to avoid analyzing the problem of operator ordering, which in this context makes it necessary to do some annoying and rather fruitless checking that there are no inconsistencies arising from operator ordering (one may check that they are not). Second, and more importantly, from the analysis on the phase space we will be able to turn directly to the analysis of special relativistic effects, by making use of the fact that what we will have at hand, would be a theory with rather unusual symplectic structure, but still defined on a decent, commuting phase space.  From now on, therefore the bracket symbol represents the Poisson bracket, and the variables (for whose I use now small letters instead of capitals), denote functions on phase space and not operators.

There are in principle many ways how one can extend the algebra (\ref{4}) along with (\ref{5}) to the whole of the phase space. However having in mind that we demand that all four components of momenta and positions $x_i$   commute among themselves
$$ [ p_\mu, p_\nu] =0, \quad [ x_i, x_j] =0, $$ this freedom is drastically reduced. For example it follows from Jacobi identity that the commutators of $\vec{p}$ with $\vec{x}$  cannot depend on $p_i$, $x_i$. One of the possible algebras (called ``Heisenberg-double'' and analyzed first in \cite{luno})
$$
  [ x_0, x_i] = \ell/\hbar \, x_i, \quad
  [ x_0, p_i] = - \ell/\hbar \, p_i,
$$ $$
  [ x_0, p_0] = 1,\quad
  [ x_i, p_j] = -\delta_{ij},$$
leads, after the procedure analogous to that I will present below, to the constraint $2\hbar =\ell\kappa$ which is incompatible with the co-product structure of the Hopf algebra, eq.~(\ref{rel}), and for this reason it is not applicable in the context of this paper.

Here I will study the algebra of the form ( a class of such algebras has been introduced in \cite{luruza})
\begin{equation}\label{6}
  [ x_0, x_i] = \ell/\hbar \, x_i,
\end{equation}
\begin{equation}\label{7}
  [ x_0, p_0] = 1,
\end{equation}
\begin{equation}\label{8}
  [ x_i, p_j] = -\delta_{ij}f(p_0),
\end{equation}
with all other brackets equal zero\footnote{One may wonder, why I put $1$ on the left hand side of (\ref{11}). The reason is that had I put a function $g(p_0)$ instead, it could be absorbed into redefinition $x_0$.}. Let us investigate the Jacobi identity for this algebra. The only non-trivial identity is of the form
$$ [p_j, [x_0, x_i]] + [x_i, [p_j, x_0]] + [x_0, [x_i, p_j]] =0,$$
from which it follows that $$\frac{f'}{f}=\frac\ell\hbar$$

\begin{equation}\label{9}
   f(p_0) = \exp\left(\frac\ell\hbar\, p_0 \right) \quad \left( = e^{-p_0/\kappa} \,\, \right). 
\end{equation}

This completes the construction of the algebra. Now we will use it to deduce the way how the positions $x$ transform under action of boosts.

\section{The invariance of algebra: transformation\\ rules for $x_i$ and $x_0$}

In this section I will study how the   positions transform under boosts. As  stressed above, the action of the rotation part of the $\kappa$-Poincar\'e algebra is not deformed, and since the  algebra presented in the preceding section were manifestly rotationally covariant, we are not going to be concerned with rotations anymore.

The procedure is rather straightforward: we use the known transformation rules for momenta (\ref{5}), the Leibniz rule, and try to discover what are the rules for transformation of $x_0$ and $x_i$, so that the algebra is covariant with respect to boosts. Consider, for example, the bracket (\ref{7}). We have
$$ [\delta_i x_0, p_0] + [x_0, \delta_i p_0] = [\delta_i x_0, p_0] =0,$$
where the vanishing of the second term follows from the fact that it  equals  $[x_0,p_i]$. It  then follows that

\begin{equation}\label{10}
  \delta_i x_0 = A(p_0;\kappa) x_i + \ldots
\end{equation}
where $\ldots$ denote terms which have vanishing bracket with $p_0$, and the function $A(p_0;\kappa)$ must go to $1$ when $p_0$ is small as compared to $\kappa$. 

By making use of the rotational symmetry one can find a most general expression for $\delta_i x_0 $ and $\delta_i x_j $, linear in $x$ and then, after straightforward but rather tedious computations, the condition that the algebra (\ref{6}--\ref{8}) is to be covariant  along with the condition
$$[\delta_k x_i, x_j] + [ x_i, \delta_k x_j]=0
$$
leads to 
$$
\delta_i x_0 = \exp\left[-\left(\frac\ell\hbar + \frac2\kappa \right)p_0 \right] \,  x_i  , 
$$
and
$$
 \delta_i x_j = \exp\left(\frac\ell\hbar\, p_0 \right)\, \delta_{ij}\, x_0  + \frac1\kappa \delta_{ij}\,\vec{x}\cdot \vec{p} - \frac1\kappa\, x_i \, p_j + \left(\frac\ell\hbar+ \frac1\kappa\right) x_j\, p_i. 
$$
Up till now I treated $\ell$ as if it was an independent parameter. Taking  $\ell   = -\frac\hbar\kappa$, one can bring this algebra into the final, simple form
\begin{equation}\label{11}
\delta_i x_0 = \exp\left(- \frac1\kappa \,p_0 \right) \,  x_i  , 
\end{equation}
and
\begin{equation}\label{12}
 \delta_i x_j = \exp\left(- \frac1\kappa \,p_0 \right)\, \delta_{ij}\, x_0  + \frac1\kappa \delta_{ij}\,\vec{x}\cdot \vec{p} - \frac1\kappa\, x_i \, p_j . 
\end{equation}
Since the deformed boost transformations (\ref{11}), (\ref{12}) are linear, one can readily find the quadratic form in $x_0$, $\vec{x}$, invariant under them. Surprisingly enough, this  is just the Minkowski distance:

\begin{equation}\label{13}
 d^2= x_0^2 - \vec{x} \,{}^2
\end{equation}
However the transformations (\ref{11}), (\ref{12}) are not just the Lorentz boosts, because the momenta  do transform together with positions. This can be seen most easily if,   by introducing the rapidity parameter $\xi$ one rewrites them in the form of differential equations (in two dimensions) 

\begin{equation}\label{14}
 \frac{d\,x_0(\xi)}{d\,\xi} = \exp\left(- \frac1\kappa \,p_0(\xi) \right) \,  x(\xi),
\end{equation}
\begin{equation}\label{15}
 \frac{d\,x(\xi)}{d\,\xi} = \exp\left(- \frac1\kappa \,p_0(\xi) \right) \,  x_0(\xi).
\end{equation}
These equations can be easily solved to give

\begin{equation}\label{16}
 x_0(\xi) = A \sinh(z(\xi)) + B \cosh(z(\xi)), \quad x(\xi) = A \cosh(z(\xi)) + B \sinh(z(\xi)),
\end{equation}
where

\begin{equation}\label{17}
  z(\xi) = \int \, d\xi \exp\left( -\frac1\kappa \,p_0(\xi) \right).
\end{equation}
Note that the deformed boosts (\ref{16}) act very similarly to those of the standard special relativity with the only difference that the rapidity parameter $\xi$ of the latter is replaced in the case at hands by the function $z(\xi)$, (\ref{17}). 
\newline

Having obtained the transformation rules, let us pause for a moment to try to find out what is their physical meaning. The first question one has to answer is as to how the presence of momenta variables in the positions transformation rules is to be interpreted. To answer this question we should recall that what we have at hands is the picture in the phase space, that is $x$ should not be interpreted as a coordinates of abstract space-time points but rather as events: ``a particle is here now''. From this perspective it is not really surprising that the way the position transforms do depend on the momentum carried by the particle. On  the contrary, the fact that such a dependence is not present in special relativity should be regarded rather an accident than a rule, and there is no reason that such property is to be present in any theory being a deformed extension of special relativity. It should be stressed that even though the transformation rules (\ref{16}) are so much similar to the standard ones that it makes possible to use virtually all the technical tricks of the standard special relativity, the conceptual picture is completely different. We will address this problem in the future publication \cite{gacjkgfut}.

Let us now return to eq.~(\ref{17}). It was shown in \cite{rbgacjkg} that (in two dimensions)

\begin{equation}\label{18}
p_0(\xi) = p^{(0)}_0+\kappa\, \log\left(1-\frac{p^{(0)}}{\kappa} (A-A\, {\cosh}(\xi)-{\sinh}(\xi))\right),
\end{equation}
where $p^{(0)}_0$ and $p^{(0)}$ are initial values of energy and momentum, respectively, and  the parameter $A$ equals
\begin{equation}\label{19}
  A=\kappa\, \sinh\left(\frac{p^{(0)}_0}{\kappa}\right)\, 
  \frac{e^{-p^{(0)}_0/\kappa}}{ p^{(0)}}+\frac{{p^{(0)}}^2}{2\kappa}
\end{equation}
Let us assume now that the particle is initially at rest ($p^{(0)}=0$), and its rest energy $p^{(0)}_0 =M$. Then
\begin{equation}\label{20}
 p_0(\xi) = M+\kappa\, \log\left(1-\alpha + \alpha\, {\cosh}(\xi)\right),\quad \alpha = \frac12 \left( 1 - e^{-2M/\kappa}\right)
\end{equation}
and therefore
\begin{equation}\label{21}
 z(\xi) = 2 e^{M/\kappa}\, \mbox{artanh}\left[e^{-M/\kappa}\, \tanh\left(\frac\xi2\right)\right].
\end{equation}
One can readily see that for small $\xi$, $z(\xi) \sim \xi$ and the transformations (\ref{16}) become the standard Lorentz boosts. For large $\xi$, on the other hand, the boosts saturate $z(\xi) \sim const$, and both $x_0(\xi)$ and $x(\xi)$ become constant (in four dimensions in the large $\xi$ limit boosts become pure rotations.)
 This behavior reminds very much the qualitative description presented in \cite{suss}.

\section{Motion of free particles and length contraction}

Having the structure of phase space (\ref{6} --\ref{8}) it is a straightforward exercise to derive equations of motion for free particles. The Poisson brackets algebra among phase space coordinates $Y_A = (x_0, x_i; p_0, p_i)$ provides us with the symplectic structure $\omega_{AB}$ and the equation of motion read
\begin{equation}\label{22}
 \frac{d\, Y_A}{d s} = \omega_{AB} \frac{\partial\, H}{\partial\,Y_B},
\end{equation}
where $s$ is the affine parameter. Similarly one can find the hamiltonian form action describing the free motion of a particle.
Both the equations and the action are by construction covariant (resp.~invariant) with respect to deformed transformation if the hamiltonian $H$, used to describe the one-particle dynamics is  invariant with respect to these transformations. It is natural therefore (and consistent with the fact that in the limit $\kappa\rightarrow\infty$ equation (\ref{22}) should become the standard equation of special relativity) to choose the hamiltonian to be equal to the Casimir (\ref{2})
$$ H = \left(2\kappa \sinh\left(\frac{p_0}{2\kappa}\right)\right)^2 - \vec{p}\,{}^2 e^{p_0/\kappa}.$$

In such  a case the action for a particle takes the form

\begin{equation}\label{22a}
  S = \int ds \, p_0 \, \frac{d}{ds} \left( x_0 + \frac1\kappa\, e^{p_0/\kappa}\, \vec{x}\cdot \vec{p} \right) - e^{p_0/\kappa}\, \vec{p} \, \frac{d}{ds} \vec{x} - H
\end{equation}

while equation of motion become

\begin{equation}\label{23}
  \frac{d\, x_i}{ds} = p_i,
\end{equation}

\begin{equation}\label{24}
 \frac{d\, x_0}{ds} = \kappa \sinh \frac{p_0}\kappa - e^{p_0/\kappa} \frac{\vec{p}\,{}^2}{2\kappa},
\end{equation}

\begin{equation}\label{25}
  \frac{d\, p_i}{ds} =\frac{d\, p_0}{ds} =0.
\end{equation}
These equations can be used to bring the action (\ref{22a}) to the ``second order'' form. I will present the study of this action in a separate paper.

Using  equations (\ref{23}), (\ref{24}) one can easily derive an expression for components of velocity of particles (in a particular reference frame)
\begin{equation}\label{26}
 V_i \equiv \frac{dx_i}{dx_0} =  \frac{dx_i/ds}{dx_0/ds} = \frac{p_i}{\kappa \sinh \frac{p_0}\kappa - e^{p_0/\kappa} \frac{\vec{p}\,{}^2}{2\kappa}}.
\end{equation}
 The first question to ask is what is velocity of massless particles. A straightforward computation of $$\vec{V}\,{}^2 = \vec{p}\,{}^2\,\left[\kappa \sinh \frac{p_0}\kappa - e^{p_0/\kappa} \frac{\vec{p}\,{}^2}{2\kappa}\right]^{-2}$$ in the massless case, i.e., when the condition $$0={\cal M}^2 = \left(2\kappa \sinh\left(\frac{p_0}{2\kappa}\right)\right)^2 - \vec{p}\,{}^2 e^{p_0/\kappa}$$ is satisfied, leads to the simple result

\begin{equation}\label{27}
 \vec{V}_{{\cal M}=0}^2 =1
\end{equation}
 Therefore the massless particles move (in space-time) with the universal speed of light. This result seems quite alarming if one recalls the expression for ${\cal C}$, (\ref{3}). This puzzle may be however solved as follows. The formula (\ref{3}) describes the group velocity of the wave, while (\ref{27}) corresponds to its phase velocity. In the standard optics these two velocities differ as a result of the dependence of the refraction index on frequency. Here the situation is quite different. The apparent discrepancy between these two expressions arises if one assumes that the relation of energy $p_0$ and frequency $\omega$ and momentum $p$ and wavelenght $\lambda$ is standard, as in special relativity. But this does not necessarily need to be  the case here. One can easily check that if one takes
\begin{equation}\label{27a}
  \omega = \kappa\sinh\left(\frac{p_0}\kappa\right) + \frac{1}{2\kappa}\, p^2 \, e^{p_0/\kappa}
\end{equation}
and
\begin{equation}\label{27b}
 \frac1\lambda = e^{p_0/\kappa}\, p
\end{equation}
and uses equation (\ref{3}) (or dispersion relation) to derive the phase velocity, the result is
\begin{equation}\label{27c}
  \frac{d\, \omega}{d\, (\lambda^{-1})} =1,
\end{equation}
in complete agreement with eq.~(\ref{27}). It should be stressed  that what was said above is just a  claim (very plausible, however), which will have to be justified by more solid arguments, for example by analyzing propagation of waves and investigating the particle--wave duality in depth. In other words, equations (\ref{27a}) -- (\ref{27c}) will have to be derived from some first principles.
 \newline
 
 Let us investigate now the behavior of massive particles. Consider a particle at rest with the initial value of energy $p_0=M$. If we boost this particle with the rapidity $\xi$, its momentum and energy will be equal \cite{rbgacjkg}
\begin{equation}\label{28}
  p(\xi) = \frac{\kappa\vartheta_0\sinh(\xi)}{\vartheta_0\cosh(\xi) +(1-\vartheta_0)},
\end{equation}
\begin{equation}\label{29}
  p_0(\xi) = M + \kappa\log\left[\vartheta_0\cosh(\xi) +(1-\vartheta_0)\right],
\end{equation}
where

\begin{equation}\label{29a}
  \vartheta_0 = e^{-M/\kappa}\, \sinh\left(\frac{M}\kappa\right) <1
\end{equation}
To obtain an expression for velocity one should substitute these formulas to (\ref{26}):

\begin{equation}\label{30}
 V_\xi = \frac{\sinh(\xi)}{\sinh\left(\frac{M}{\kappa}\right) +\cosh\left(\frac{M}{\kappa}\right)\, \cosh(\xi) }.
\end{equation}
Observe that in this way one obtains a measure of the relative velocities of two observers $V$ as a function of the rapidity parameter $\xi$. Let us analyze this expression in the limit of  boosts going to infinity. One gets
\begin{equation}\label{31}
 V_\infty = \cosh\left(\frac M\kappa\right)^{-1}.
\end{equation}
This means that even applying infinite boost, contrary to special relativity, one cannot make massive particle moving with the universal speed of light.
\newline

Let us now turn to the last problem, namely the issue of time dilatation and length contraction. By inspecting the formulas (\ref{16}) and using the standard reasoning of special relativity one would claim that if the ``rest'' time difference between events at the same point equals $\tau$, the moving observer will attribute it the time difference $\tau\, \cosh(\xi)$; similarly, if the ``rest'' distance equals $l$, a moving observer will attribute it the distance $\frac{l}{\cosh(\xi)}$. Let us now show that this expectation is indeed correct. To do so I will not make use of any kind of ``time of flight'' thought experiments, because they require introduction of mirrors, clocks and other devices that are conceptually external to the structure of the theory.

Consider time dilatation. Take a particle at rest described by a solution of equations of motion with momentum $p=0$:
\begin{equation}\label{32a}
  x[s] =0, \quad x_0[s] = s\kappa\sinh\left(\frac{p_0}{\kappa}\right)
\end{equation}
Then\footnote{Observe that in order to get this result I make use of the metric structure of space-time. It is important to note that without this structure there is no sensible way of talking about distances.}
\begin{equation}\label{32b}
 \tau = x_0[s_2] - x_0[s_1] = (s_2 - s_1)\, \kappa\sinh\left(\frac{p_0}{\kappa}\right)
\end{equation}
Let us now boost the particle with rapidity $\xi$. We have
\begin{equation}\label{32c}
  x[s;\xi] =s\kappa\sinh\left(\frac{p_0}{\kappa}\right)\, \sinh(z(\xi)), \quad x_0[s;\xi] = s\kappa\sinh\left(\frac{p_0}{\kappa}\right)\, \cosh(z(\xi))
\end{equation}
and thus
\begin{equation}\label{32d}
  \tau' = x_0[s_2;\xi] - x_0[s_1;\xi] = \tau \, \cosh(z(\xi))
\end{equation}
which is indeed the expected result.

To derive the contraction formula, consider two particles (of  masses $M_1$ and $M_2$) at rest in one coordinate system: the first in position $x^{(1)} =0$ and the second in position $x^{(2)}=l$. Solving their equation of motion one gets
\begin{equation}\label{32}
  \begin{array}{ll}
   x^{(1)}[s] = 0 & x^{(2)}[\bar s]=l \\
    x_0^{(1)}[s] = s \kappa \sinh\left(\frac{M_1}\kappa\right)&  x_0^{(2)}[\bar s] = \bar s \kappa \sinh\left(\frac{M_2}\kappa\right)\
  \end{array}
\end{equation}
If we now boost the particle $(1)$ its motion will be described by
\begin{eqnarray}
x^{(1)}[\xi;s] &=& s \kappa \sinh\left(\frac{M_1}\kappa\right) \sinh(z(\xi))\nonumber\\
x^{(1)}_0[\xi;s] &=& s \kappa \sinh\left(\frac{M_1}\kappa\right) \cosh(z(\xi)),\label{33}
\end{eqnarray}
while for particle (2), one has
\begin{eqnarray}
x^{(2)}[\xi;\bar s] &=& l\cosh(z(\xi)) + \bar s \kappa \sinh\left(\frac{M_2}\kappa\right) \sinh(z(\xi))\nonumber\\
x^{(2)}_0[\xi;\bar s] &=& l\sinh(z(\xi)) + \bar s \kappa \sinh\left(\frac{M_2}\kappa\right) \cosh(z(\xi)),\label{34}
\end{eqnarray}
The distance between the particles measured by boosted observer $l'$ will be equal to the difference between $x^{(2)}[\xi;\bar s]$ and $x^{(1)}[\xi; s]$ with $s$ and $\bar s$ such that $x^{(1)}_0[\xi;s] =  x^{(2)}_0[\xi;\bar s]$. The simple computation shows that

\begin{equation}\label{35}
 l' = l\cosh(z(\xi)) - l \frac{\sinh(z(\xi))^2}{\cosh(z(\xi))} = l\frac{1}{\cosh(z(\xi))}
\end{equation}
which is exactly the expected result. Observe that since the function $z(\xi)$ goes to the constant limit when $\xi\rightarrow\infty$, the fraction $l/l'$ is, contrary to special relativity, limited from above. 

\section{Conclusions and outlook}

I start with collecting the main results presented above:

\begin{enumerate}

\item The starting point of the construction is an algebra consisting of the quantum $\kappa$-Poincar\'e algebra together with eqs.~(\ref{6}) (being in one-to-one correspondence with the co-product structure of the quantum $\kappa$-Poincar\'e algebra) and (\ref{7}), (\ref{8}). The assumption that this algebra is to be covariant with respect to action of the boosts leads to the infinitesimal and finite transformation rules for positions (\ref{11}), (\ref{12}) and (\ref{16}), respectively. One of the most important conclusions is that the boosts saturate as in the momentum space, \cite{rbgacjkg}.
\item It is proved that the metric structure of so constructed ``space-time of events'' possesses the Minkowski distance as an invariant quadratic form. This means that this space is $R^4$ equipped with Minkowski metric (contrary to the energy-momentum space, for whose the invariant form is the Casimir, (\ref{2}).
\item Comparing the formulas for velocities ${\cal C}$ (\ref{3}) and the one following the Minkowski structure of the space-time of events I propose the relation between energy and momentum on the one side and frequency and inverse wavelength, on the other (\ref{27a}), (\ref{27b}).
\item The action for and equation of motion of massive and massless point particles and formulas describing time dilatation and length contraction are derived. 
\end{enumerate}

This paper is first attempt to describe the space-time picture emerging from the $\kappa$-Poincar\'e framework and, of course, there is a lot of open problem which will have to be addressed. First of all one will have to try to understand if the claimed relation of point 3.~above is indeed correct and if so what are its physical and conceptual consequences. It seems that this relation together with the algebra (\ref{6})--(\ref{8}) may have something to do with the possible appearance of a new, observer-independent length scale (recall that here we deal with two such scales: the universal speed of light $c=1$ and the mass scale $\kappa$; see \cite{gac1}, \cite{gac2}, \cite{jkgminl}, \cite{rbgacjkg}).

Also, the results obtained in this paper may provide a starting point to an attempt to construct a curved space theory based on $\kappa$-Poincar\'e algebra. The idea would be to give this algebra the status analogous to that played by the standard Poincar\'e algebra in the general relativity and to try to gauge it. It is not clear if it would be possible (or desirable) to keep diffeomorphism invariance. It seems also interesting to try to develop a theory of early cosmology similar to the one proposed in \cite{stjo}, which would take into account the results presented above and to check out if such a theory would be capable of solving  puzzles of the standard cosmological model.

Last but not least one should attempt to develop an overall conceptual framework which would make it possible to provide a proper physical interpretation and understanding of the results presented in this paper. This issue will be addressed in \cite{gacjkgfut}.

\section*{Acknowledgement} I would like to thank  Giovanni Amelino-Camelia for his hospitality in Rome, where my work on this project started  and many illuminating discussions. Thanks are due to him, Ted Jacobson and Jerzy Lukierski for their valuable comments on the draft of this paper. This research is partially supported by the   KBN grant 5PO3B05620.

\end{document}